\documentclass[aps,prl,twocolumn,reprint]{revtex4-2}
\usepackage[english]{babel}
\usepackage[utf8]{inputenc}
\usepackage[colorinlistoftodos, color=green!40, prependcaption]{todonotes}
\usepackage{amsthm}
\usepackage{amsmath}
\usepackage{amssymb}
\usepackage{mathtools}
\usepackage{physics}
\usepackage{xcolor}
\usepackage{graphicx}
\usepackage[left=20mm,right=20mm,top=25mm,columnsep=15pt]{geometry} 
\usepackage{comment}
\usepackage{adjustbox}
\usepackage{placeins}
\usepackage[T1]{fontenc}
\usepackage{lipsum}
\usepackage{csquotes}
\usepackage{hyperref}
\usepackage{CJKutf8}
\newcommand{\tg}[1]{\textcolor{blue}{#1}} 

\usepackage{tikz}

\begin{document}

\title{Isosbestic points in time resolved SAXS: from spectroscopic analogy to model free structural markers during colloidal gelation}
\author{Alain Gibaud,$^{1,2}$ Wilbert J. Smit,$^{3}$ Safa Jamali,$^{4}$ and Thomas Gibaud$^{3,5}$}

\email[Corresponding authors:~]{alain.gibaud@univ-lemans.fr, thomas.gibaud@ens-lyon.fr}

\affiliation{%
$^1$ IMMM, UMR 6283 CNRS, Le Mans Université, 72085 Le Mans, France \\
$^2$ SWUST, Sichuan, Mianyang, Fucheng District,
\begin{CJK}{UTF8}{gbsn}青龙大道中段\end{CJK}59\begin{CJK}{UTF8}{gbsn}号, 邮政编码\end{CJK}: 621010, China \\
$^3$ ENSL, CNRS, Laboratoire de Physique, F-69342 Lyon, France \\
$^4$ Department of Mechanical and Industrial Engineering, Northeastern University, Boston, MA 02115, USA\\
$^5$ Department of Polymer Engineering, IPC, University of Minho, Guimarães, 4804-533 Portugal}


\date{\today}

\begin{abstract}
Gelation is the transition from a fluid state into a system-spanning, out of equilibrim soft-solid network through a hierarchical process that couples local particle interactions to mesoscopic clustering and global connectivity. In time-resolved small-angle X-ray scattering (SAXS), isosbestic points—scattering wavevectors where scattering intensity remains invariant—emerge during this transformation, yet their physical meaning has remained unclear. Here, we show that two isosbestic points, $q_1$ and $q_2$, observed during salt-induced gelation of Ludox colloids, reflect fundamental structural constraints rather than a two-species interconversion. The high-$q$ point $q_2$ is a universal geometric marker, determined by particle contact distances, while the low-$q$ point $q_1$ arises from Porod invariant conservation and separates rapidly arrested local clusters from the growing mesoscopic network. By decomposing the Porod invariant across the reciprocal-space regions defined by these points, we define a dimensionless parameter, $\Phi(t/t_g)$, whose sigmoidal evolution provides a simple, model-free, scale-resolved fingerprint of gelation. Together with the combined evolution of $S(q_{\min},t)$ and $S(q \rightarrow 0,t)$, these results establish a quantitative model free framework linking local structuring, global connectivity, and scattering signatures, clarifying the role of isosbestic points in soft-matter transformations.
\end{abstract}

\maketitle

\section{Introduction}
Gelation belongs to a broad class of non-equilibrium transformations in soft matter, in which particles in fluid state self-assemble into a space-spanning network to form a soft-solid state. This process is inherently multiscale: local particle interactions progressively give rise to mesoscopic clusters, which ultimately connect into a system-spanning structure~\cite{poon1997,gibaud2012,bauland2024,bauland2026}. Capturing this hierarchical and continuous reorganization remains a central challenge, whether one works in real space using microscopy or in reciprocal space using scattering~\cite{royall2021,macdonald2024}.

Focusing on scattering techniques, the gel structure is encoded in reciprocal space, and the loss of phase inherent to classical scattering experiment implies that only spatial correlations—rather than absolute particle positions—are accessible. Consequently, the measured intensity reflects pair correlations through the structure factor, making the interpretation of structural evolution intrinsically indirect. Furthermore, modeling remains challenging in this non-equilibrium context. Classical equilibrium approaches based on liquid-state theory~\cite{pontoni2003,sztucki2006} are not always applicable, especially when the attraction between colloids is too strong. Geometric models, such as mass fractal or unified scattering descriptions~\cite{freltoft1986,teixeira1988,beaucage1995}, rely on simplified assumptions and often capture only limited regions of the scattering curve. This motivates the search for robust, model-independent observables that directly reflect structural evolution and can complement model-dependent approaches.

\textcolor{black}{A further complication is that the mechanically relevant structure of a colloidal gel is not necessarily captured by a single geometric descriptor. Recent particle-resolved and network-based studies have shown that colloidal gelation involves the emergence, aging, and restructuring of a particulate network whose rheological response depends not only on cluster size or volume fraction, but also on bond lifetime, local connectivity, load-bearing pathways, and the history by which the network was formed ~\cite{Mangal2024, Nabizadeh2021, Bantawa2023, Nabizadeh2024}. In this sense, gelation is not only a problem of identifying a characteristic length scale, but also one of determining how structural correlations become mechanically connected across scales. This distinction is particularly important because different microstructures may share similar conventional scattering or connectivity descriptors while retaining different mechanical memories. These considerations motivate structural observables that are not tied to a particular fitting model, but that can nevertheless separate local aggregation, mesoscale arrest, and global network formation.}

In this context, invariant features in experimental signals provide particularly appealing markers. An invariant (or isosbestic) point corresponds to a condition where the measured response remains unchanged despite ongoing structural or compositional evolution, typically reflecting an underlying conservation constraint. A well-known realization is found in absorption spectroscopy, where an isosbestic point corresponds to a wavelength at which the total absorbance remains constant during a transformation~\cite{geissler2005}. In such cases, the signal can often be described as a linear superposition of contributions from distinct molecular species at fixed total concentration, leading to a crossing point independent of their relative populations. Classical examples include acid–base equilibria, such as bromothymol blue, where the absorption spectra intersect at a well-defined wavelength. 
In time resolved SAXS measurement associated with out of equilibrium self-assembly, single to multiple Isosbestic points have also been reported, where intensity profiles $I(q,t)$ intersect at specific wavevectors. Such features have been often interpreted within a two-state framework~\cite{nicolai2006, goodell2008, sauter2016, tyburski2024}. For instance, Nicolai \emph{et al.}~\cite{nicolai2006} analyzed protein aggregation using a linear superposition of scattering contributions from native and denatured states, successfully reproducing the observed isosbesting point. Similar observations have been reported in gel-forming systems, including protein cross-linking~\cite{kaieda2014}, bicontinuous emulsion gelation~\cite{alting2025} and colloidal gelation~\cite{wyss2004}, although the origin of these features was not further analyzed.
The key difference with spectroscopy lies in the nature of the measured signal. While absorption probes additive molecular responses, scattering intensities arise from interference between all scatterers and therefore encode spatial correlations. As a consequence, the total intensity cannot, in general, be decomposed into a simple linear combination of independent contributions without restrictive assumptions, such as spatial separation or statistical independence. These conditions are unlikely to be satisfied in continuously evolving systems such as colloidal gels, where structure develops across overlapping length scales. 
This leads to several open questions. When isosbestic points are observed in SAXS during gelation, do they genuinely indicate a two-state structural transformation, or do they emerge more generally from conservation constraints and the redistribution of correlations across length scales? Can these invariant points be associated with specific structural features, spanning from local particle contacts to mesoscopic clusters? More generally, can they be used to construct simple, model-free descriptors of the gelation pathway?

\textcolor{black}{These questions are closely related to a broader issue in colloidal gel physics as well: how reciprocal-space signatures should be connected to the evolving real-space network ~\cite{PanizHaghighi2025}. In particle-resolved descriptions, gelation proceeds through the formation of bonds, clusters, strands, pores, and eventually a percolated backbone; however, these structural motifs are distributed over a broad range of length scales and need not evolve synchronously. For example, temporal control of gelation in multicomponent colloidal systems can generate markedly different morphologies—from mixed networks to core–shell or double-network-like structures—even when the particle-level interactions are fixed ~\cite{Kaltashov2026}. Similarly, under deformation, gels with apparently convergent structural descriptors can retain formation-history-dependent yielding, indicating that structural convergence does not necessarily imply mechanical equivalence. These examples emphasize the need for scale-resolved markers that can distinguish local arrest from global connectivity without assuming a unique real-space morphology.}

Here, we address these questions using time-resolved SAXS measurements of salt-induced gelation in Ludox colloidal dispersions across three particle sizes, combined with simultaneous rheology. Because the particle form factor remains constant throughout the transformation, the measured intensity directly reflects the evolution of the structure factor, providing a clean framework to analyze invariant features.
We systematically observe two isosbestic points. The high-$q$ point is a true isosbestic point, whereas the low-$q$ point is only pseudo-isosbestic, exhibiting a small but systematic drift during the early stages of gelation. 
We first test whether these isosbestic points can be explained by a simple two-state model, based on a linear superposition of liquid and fully gelled states. While this model is insufficient to fully capture the data, an extended formulation incorporating cross-interference terms provides a remarkably accurate quantitative description  over the full time evolution. Small deviations from the model, particularly at low $q$, highlight the limits of this approach indicating that the model’s success is only effective, reflecting continuous structural evolution rather than true species interconversion.
Second we use those two isosbestic points to partition reciprocal space into three distinct regimes associated from low to high-$q$ with large-scale density fluctuations, cluster formation, and single-particle structure. We show that these points have different physical origins: the high-$q$ point is associated with a geometric constraint linked to the first coordination shell, while the low-$q$ point emerges from the conservation of the Porod invariant, acting as a pivot between local aggregation and network formation. 
Finally, building on this interpretation, we decompose the Porod invariant across the $q$-regions defined by the isosbestic points, yielding a scale-resolved perspective of density redistribution and introducing a dimensionless parameter that quantifies gelation. Moreover, the combined temporal evolution of the structure factor minimum, $S(q_{\min}, t)$, and the low-$q$ limit, $S(q \rightarrow 0, t)$, provides a direct, model-free signature of gelation. This analysis reveals a decoupling between the rapid formation of local structures and the slower establishment of system-spanning connectivity.
Together, these results establish simple and quantitative fingerprints of colloidal gelation and clarify the physical significance of isosbestic points in scattering measurements.
\section{Materials and Methods}
\label{sec:methods}
\subsection{Ludox colloids and salt induced gelation}
In the present work, we follow the gelation of aqueous dispersions of Ludox silica nanoparticles. Gelation is induced by the addition of sodium chloride, which screens the electrostatic repulsion between the negatively charged Ludox-particles and triggers aggregation~\cite{rueb1998, trompette2004,smit2025}. 
Three colloidal gels were prepared using Ludox particles of different radii $r_0$ and volume fractions $\phi$.
The volume fraction $\phi$ corresponding to a weight fraction $c_w$ is $\phi = \frac{c_w/\rho_s}{c_w/\rho_s + (1-c_w)/\rho_w}$ with density for ludox $\rho_s = 2.0\ \text{g/cm}^3$  and water $\rho_s = 1.0\ \text{g/cm}^3$.
We use SM particles with $r_{0,\mathrm{SM}} = 5.3\,\mathrm{nm}$, $\phi_{\mathrm{SM}} = 2.5\%$, and $[\mathrm{NaCl}] = 500\,\mathrm{mM}$; HS particles with $r_{0,\mathrm{HS}} = 7.8\,\mathrm{nm}$, $\phi_{\mathrm{HS}} = 4.8\%$, and $[\mathrm{NaCl}] = 610\,\mathrm{mM}$; and TM particles with $r_{0,\mathrm{TM}} = 12.9\,\mathrm{nm}$, $\phi_{\mathrm{TM}} = 11.1\%$, and $[\mathrm{NaCl}] = 700\,\mathrm{mM}$. See appendix.A for more information. 


\subsection{Rheo-SAXS experiment}
SAXS and rheology measurements were performed simultaneously. SAXS experiments were conducted at the ID02 beamline of the European Synchrotron Radiation Facility (ESRF) in Grenoble, France~\cite{Narayanan2022}. The incident X-ray beam, with a wavelength of approximately 0.1~nm ($E=12.23$ keV), was collimated to 50~$\mu$m vertically and 100~$\mu$m horizontally. Two-dimensional scattering patterns were recorded over time using an Eiger2 4M pixel array detector. The scattering intensity \(I(q,t)\) was obtained by subtracting the background scattering of the solvent from the two-dimensional profiles of the dispersion. The scattering remained isotropic throughout the study, and an azimuthal average was performed to produce one-dimensional \(I(q,t)\) profiles. The dispersion was loaded into a Haake RS6000 (Thermo Scientific) rheometer equipped with a Couette geometry consisting of concentric polycarbonate cylinders (inner diameter 20 mm, outer diameter 22 mm, height 40 mm). The X-ray beam intersected the Couette cell along the rotation axis of the rheometer (radial configuration).


\subsection{SAXS data analysis}
\noindent\textbf{Isosbestic point determination -- }
In SAXS, an isosbestic point is defined as a wavevector $q_{\rm iso}$ at which multiple scattering curves intersect, such that for any two distinct times $t_i$ and $t_j$, $I(q_{\rm iso},t_i)=I(q_{\rm iso},t_j)$. 
In the code, all pairwise differences between scattering curves are computed, and sign changes are used to detect crossings. Linear interpolation is then applied to estimate the precise intersection for each pair of curves. 
We recall that $q_1(t)$ and $q_2(t)$ are the low-$q$ and high-$q$ isosbestic positions, with corresponding intensities $I(q_1,t)$ and $I(q_2,t)$. 
The error bars on $q_i(t)$ and $I(q_i,t)$ are defined as the standard deviation of all pair intersections at a given time $t$.


\vspace{2mm}
\noindent\textbf{Measurement of the form factor $P(q)$ and determination of $S(q)$ -- } The form factor $P(q)$ of the Ludox particles was measured in Milli-Q water at low volume fraction ($\phi = 0.2\%$) to minimize interparticle strutural correlation and with no added salt to avoid aggregation.  A polydisperse sphere model with a Schultz size distribution was used to fit the data. The data is displayed in Appendix.A. The resulting fitted $P(q)$ provides a smooth and physically consistent description of the single-particle scattering over the full $q$-range. As a results the scattering intensity can be decomposed as follow: $I(q,t)=\phi( r_e\Delta \rho)^2 P(q) S(q,t)$ where $r_e= 2.871\times 10^{-6}$~nm is the classical electron radius, $\Delta\rho=318$~e/nm$^3$is the electron density scattering contrast between the ludox and the aqueous background, and $S(q,t)$ is the time dependant structure factor.
As the experimental form factor remains affected by noise, especially at high $q$, we opted to extract the structure factor via $S(q,t) \propto   I(q,t)/P(q)$, where $P(q)$ is the fitted form factor.

\vspace{2mm}
\noindent\textbf{Calculation of $g(r,t)$ from $S(q,t)$ -- }
The radial distribution function $g(r)$ was obtained from the structure 
factor $S(q,t)$ via sine Fourier transformation:
\begin{equation}
g(r,t) = 1 + \frac{1}{2\pi^2 n r} \int_{q_{\min}}^{q_{\max}} q\,W(q)\,[S(q,t) - 1]\,\sin(qr)\,\mathrm{d}q,
\end{equation}
where $n = \phi/(4\pi r_0^3/3)$ is the particle number density and $W(q)$ is a 
window function applied to suppress truncation ripples arising from the finite 
upper integration limit $q_{\max}$~\cite{soper2012}. Without modification, the 
abrupt truncation of $S(q)$ at $q_{\max}$ introduces spurious oscillations of 
frequency $\sim\pi/q_{\max}$ in $g(r)$, which can obscure the structure of the 
coordination shells. We use the Lorch modification function $W(q) = 
\sin(\pi q/q_{\max})/(\pi q/q_{\max})$, which effectively suppresses these 
truncation oscillations at the cost of a modest broadening of the $g(r)$ peaks.

\section{Results}
\label{sec:results}
\begin{figure*}
\centering
\includegraphics[width=1.0\textwidth]{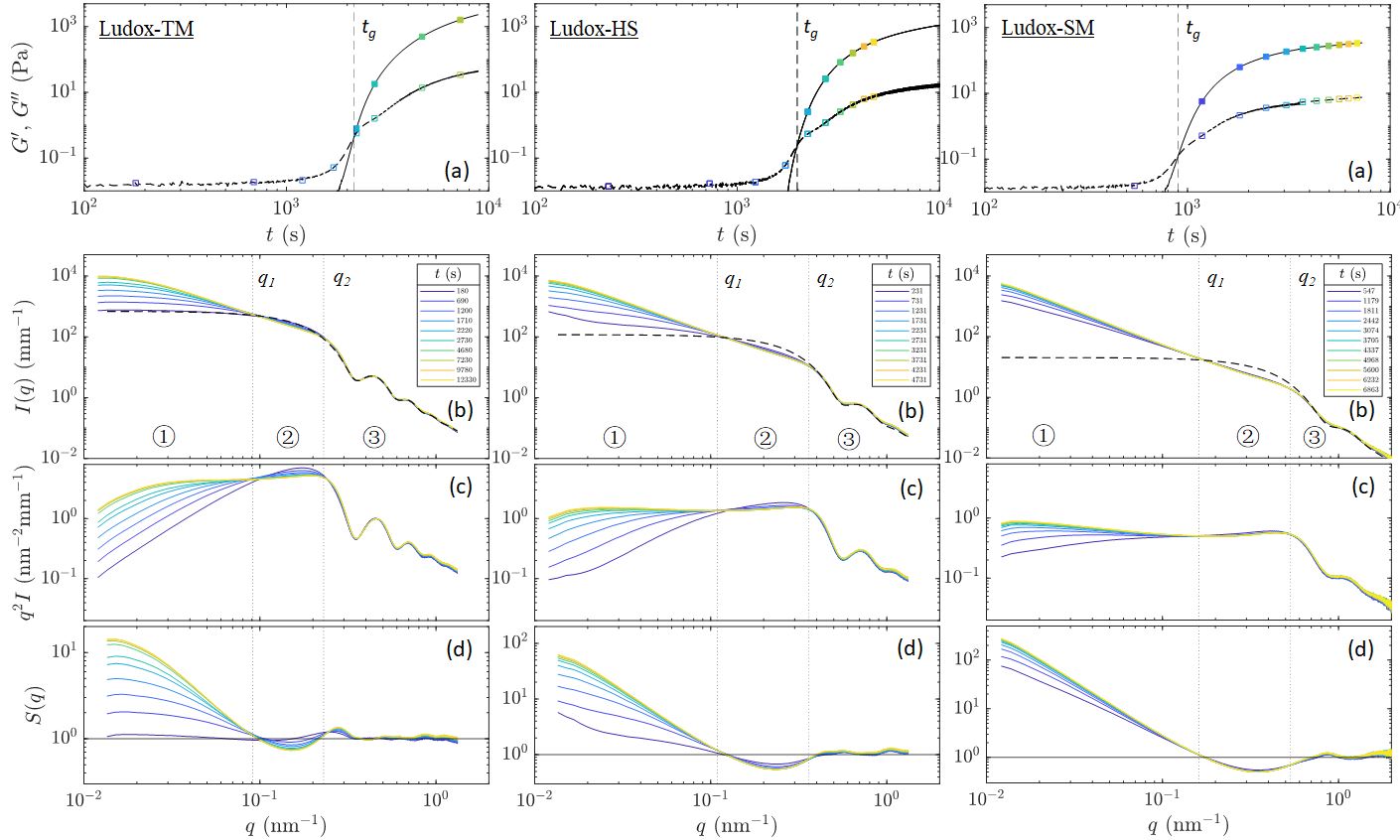}
    \caption{Gelation of ludox TM, HS and SM, rheology and scattering. (a) Time evolution of the viscous $G''$ (open symbols) and elastic $G'$ (full symbol) moduli measured under small amplitude oscillatory strain at $f=1$~Hz and $\gamma=1$~\%. The gelation time is $t_g$ defined such that $G'(t_g)=G''(t_g)$ and indicated by the vertical dash line. (b) Time evolution of the scattering intensity $I(q,t)$ as function of the norm of the scattering wave vector $q$. The color codes for time. The dash line representes the form factor rescaled. (c) Kratky representation: $q^2I(q,t)$ verus $q$. (d) Structure factor $S(q,t)$. The vertical dotted lines in (b-d) represent the position of the isosbestic point at low-$q$ ($q_1$) and high-$q$ ($q_2$) separating the scattering data into three regions noted \textcircled{1}, \textcircled{2} and \textcircled{3}. 
    }
        \label{fig:iq}
\end{figure*}

\begin{figure*}
\centering
\includegraphics[width=1\textwidth]{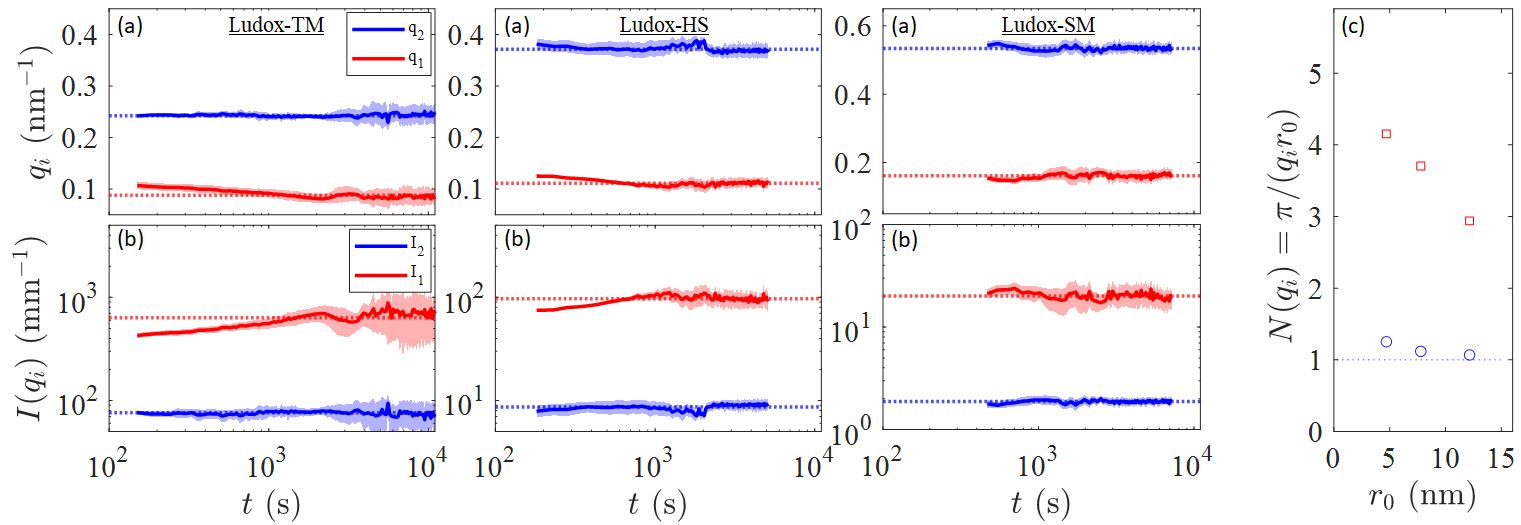}
    \caption{Isosbestic points for the TM, HS, and SM Ludox dispersions. 
(a) Evolution of the characteristic isosbestic wavevectors $q_1$ (red) and $q_2$ (blue) as a function of time $t$. 
(b) Corresponding scattering  isosbestics intensities $I_1$ (red) and $I_2$ (blue) versus time. 
The solid lines represent the average intersection obtained from pairwise curve crossings at a given time $t$. 
The dashed lines indicate the time-averaged values. 
The shaded regions around each line represent the standard deviation at each time. (c) Evolution of $N(q_i)=\pi/(q_i r_0)$ for $i=1$ ($\square$) and $i=2$ ($\circ$) as a function of the particle radius $r_0$ for the TM, HS, and SM Ludox dispersions. The red horizontal line indicates the reference value $1$. 
    }
        \label{fig:iso}
\end{figure*}
\subsection{isosbestic points in SAXS data during gelation of the ludox dispersions}
The data associated with the gelation of ludox TM, HS and SM are shown in Fig.~\ref{fig:iq}. We present the concurrent time evolution of the viscoelastic moduli (Fig.~\ref{fig:iq}(a))---the elastic modulus $G'$ and the viscous modulus $G''$ measured under small-amplitude oscillatory shear ($f = 1\,\mathrm{Hz}$, $\gamma = 1\%$)---together with the SAXS intensity $I(q,t)$, where $q$ denotes the magnitude of the scattering vector (Fig.~\ref{fig:iq}(b)). Rheological measurements show that at short times the dispersion behaves as a fluid with $G'' > G'$, while at longer times the system evolves into a soft solid characterized by a dominant elastic response ($G' > G''$). The gelation time $t_g$ is defined as the crossover time at which $G' = G''$. 
The SAXS data reveal a progressive increase of the low-$q$ scattering intensity, indicative of aggregation and the formation of fractal clusters characterized by a correlation length $\xi$ and a fractal dimension $d_f$ and typical of fractal gels~\cite{kolb1983,meakin1983}.

For the Ludox dispersions, as shown in Fig. ~\ref{fig:iq}(b), two well-defined isosbestic points are observed in the SAXS data at low and high wavevectors, denoted $q_1$ and $q_2$, respectively. In Fig.~\ref{fig:iso}, we quantitatively verify the isosbestic nature of these points by plotting the coordinates of all pairwise intersections of $I(q,t)$ over time. We find that  $(q_2, I_2)$ behaves as a true isosbestic point, with coordinates remaining constant within a relative uncertainty of about 1--2\%. For 
$(q_1, I_1)$, the relative uncertainty is slightly larger, around 5--10\%, due to a small decrease in $q_1$ accompanied by a slight increase in $I_1$ at early times compared to their time-averaged values therefore making $(q_1, I_1)$ a pseudo-isosbesting point rather than a true isosbestic point like  $(q_2, I_2)$.

Analysis of the structure factor $S(q)$ shows that the isosbestic points lie on either side of the first minimum, which remains invariant over time. More specifically, we consistently find $S(q_1)=1$, while $q_2$ corresponds to a value slightly above or below unity, as reported in Appendix.A Tab.~\ref{tab:ludox}. These observations indicate that the isosbestic points correspond to characteristic length scales at which the scattering intensity remains unchanged during gelation, and where density correlations do not contribute to the scattering. They therefore act as pivot wavevectors separating three distinct regimes of structural reorganization with distinct temporal behaviors (Fig.~\ref{fig:iq}).
In region \textcircled{1} ($q < q_1$), the strong increase in forward scattering reflects the growth and aggregation of clusters at mesoscopic length scales. 
In region \textcircled{2} ($q_1 < q < q_2$), a progressive deepening of the first minimum of $S(q)$ is observed at fixed scattering wave vector $q_{min}$. This deepening does \emph{not} indicate an increase in attraction strength—the interaction potential remains constant once salt is added and screening is established. Rather, it reflects the evolution of many-body correlations as particles aggregate via diffusive encounters under constant intermolecular forces, leading to increasingly well-defined coordination shells and short-range structural order within the forming gel network. 
In region \textcircled{3} ($q > q_2$), the scattering profile is nearly time-invariant and is dominated by the single-particle form factor $P(q)$, indicating that the internal structure of the particles remains unaffected during gelation.

\subsection{Revisiting the two-state model}

We revisit the applicability of a two-state description to the scattering data. In the context of salt-induced colloidal gelation, the most natural identification of the two states is the initial colloidal dispersion and the final gel network. Accordingly, following~\cite{sauter2016}, we define state~1 as the fluid dispersion at $t \ll t_{\rm gel}$, characterised by weak interparticle correlations and a structure factor close to unity, and state~2 as the fully developed gel at $t \gg t_{\rm gel}$, characterised by strong 
low-$q$ scattering and a well-developed first coordination shell. For the TM-dispersions, the corresponding basis functions $I_1(q) = I(q, t_{\rm start})$ and $I_2(q) = I(q, t_{\rm end})$ are taken directly from the first and last measured frames of the time series.

\begin{figure}
\centering
\includegraphics[width=0.471\textwidth]{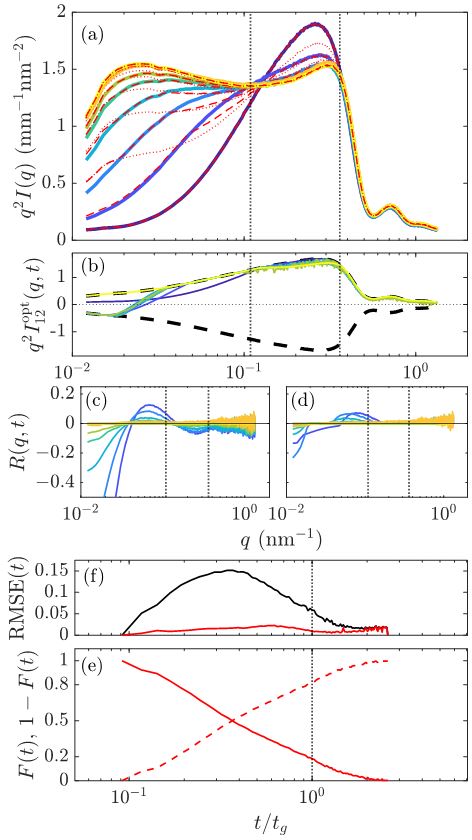}
\caption{Two-state model analysis of time-resolved ludox-TM dispersion scattering data using State-1 as the fluid state at $t \ll t_g$ and State-2 as the gel state at $t \gg t_g$. 
{(a)} Time-resolved Kratky representation of the scattering intensity versus $q$. The solid line shows experimental data, the dotted line shows the best fit using the linear model (Eq.~\ref{eq:2state}), and the dashed line shows the best fit using the full model (Eq.~\ref{eq:2state_b}). Colors code for time as in Fig.~\ref{fig:iq}.
{(b)} Kratky representation of the time evolution of the cross term (colored lines). The dashed lines indicate the upper and lower bounds according to the Cauchy--Schwarz inequality. 
{(c)} Residuals of the linear model fit. 
{(d)} Residuals of the full model fit. 
{(e)} Time-resolved $q$-averaged RMSE for the linear (black line) and full (red line) models as a function of $t/t_g$.
{(e)} Fractional contributions of the two states for the full model: $F$ (solid line) and $1-F$ (dashed line) versus $t/t_g$.
}
\label{fig:twostateb}
\end{figure}

In its simplest form, the two-state model assumes that the scattering intensity can be written as a linear combination of two limiting states,
\begin{equation}
I(q,t) = F(t)\,I_1(q) + \bigl(1 - F(t)\bigr)\,I_2(q),
\label{eq:2state}
\end{equation}
where $I_1$ and $I_2$ correspond to the intensity of each state in their pure form.
This expression neglects the cross-interference terms between the two contributions. Such an approximation is valid only if the two states are spatially independent or uncorrelated—a condition fulfilled in~\cite{nicolai2006,sauter2016}, but generally not expected especially in the case of a continuously evolving gel network.

We now consider a more general formulation of the two state model explicitly including the cross-interference term. Indeed, writing the total electron density as $\rho(\mathbf{r},t) = F(t)\,\rho_1(\mathbf{r}) + [1-F(t)]\,\rho_2(\mathbf{r})$ and expanding $I(q,t) = |\tilde{\rho}(q,t)|^2/V$ yields the full quadratic expression, 
\begin{equation}
\begin{split}
I(q,t) = {} & F(t)^2 I_1(q) + \bigl(1 - F(t)\bigr)^2 I_2(q) \\
            & + 2F(t)\bigl(1 - F(t)\bigr) I_{12}(q,t),
\end{split}
\label{eq:2state_b}
\end{equation}
where $I_{12}(q,t)$ represents the cross-correlation between the two states. This term is constrained by the Cauchy--Schwarz inequality,
\begin{equation}
|I_{12}(q,t)| \le \sqrt{I_1(q)\,I_2(q)}.
\label{eq:2state_CS}
\end{equation}

The linear superposition model of Eq.~\ref{eq:2state} is recovered from Eq.~\ref{eq:2state_b} when the cross term can be neglected, which requires the absence of correlation between states 1 and 2, or that the scattering of one state far exceeds the other in intensity.

The model parameters $F(t)$ and $I_{12}(q,t)$ are determined by minimizing the weighted least-squares of the intensity $\chi^2(t) = \sum_q W(q)\,\bigl[I(q,t) - I_{\mathrm{model}}(q,t)\bigr]^2$,
where $W(q) = 1/{\langle  I(q)\rangle_t^2},$ is a weighting function chosen to uniformly balance contributions across the $q$-range.
The quality of the fit is quantified through the relative residual $R(q,t) = \frac{I(q,t) - I_{\mathrm{model}}(q,t)}{I(q,t)}$, and its relative root-mean-square error averaged over $q$: $\mathrm{RMSE}(t) = \sqrt{\left\langle R(q,t)^2 \right\rangle_q}$.

The linear model provides a poor description of the data. As shown in Fig.~\ref{fig:twostateb}(a), the reconstructed curves deviate significantly from the experimental profiles. This is reflected in large residuals (Fig.~\ref{fig:twostateb}(c)) and a pronounced peak in the RMSE around $t/t_g \approx 0.35$, reaching values up to $15\%$ (Fig.~\ref{fig:twostateb}(f)). These discrepancies indicate that for the TM-gelation, neglecting cross-interference terms is not justified.

In contrast, the full model (Eq.~\ref{eq:2state_b}) yields a much improved description. The agreement with the data is significantly better (Fig.~\ref{fig:twostateb}(a)), with residuals reduced over most of the $q$-range (Fig.~\ref{fig:twostateb}(d)). The RMSE remains below $2\%$ over the entire time evolution (Fig.~\ref{fig:twostateb}(f)), demonstrating that the inclusion of cross terms captures the dominant features of the scattering signal.
The behavior of the cross-term $I_{12}(q,t)$ provides further insight into the evolution of spatial correlations during gelation (Fig.~\ref{fig:twostateb}(b)). Over most of the $q$-range and at late times, $I_{12}(q,t)$ approaches the upper Cauchy--Schwarz bound, indicating that the two contributions become strongly positively correlated. This reflects the fact that, as gelation proceeds, the system organizes into a single, space-spanning network where density fluctuations at a given length scale are no longer independent but instead governed by a common underlying structure.
In contrast, at early and intermediate times, and for low wavevectors ($q \lesssim 0.02$~nm$^{-1}$), $I_{12}(q,t)$ approaches the lower bound. This regime corresponds to large length scales, where the system is still evolving from a dispersed state toward a connected network. The proximity to the lower bound indicates weak or even effectively anti-correlated contributions, consistent with the presence of spatial heterogeneities such as isolated clusters and growing voids that are not yet part of a coherent structure.
Taken together, this evolution of $I_{12}(q,t)$ from the lower to the upper Cauchy--Schwarz bound provides an indirect signature of the progressive build-up of long-range correlations and the emergence of a mechanically connected gel network.

Overall, the full two-state model provides a good quantitative description of the data, with an RMSE below $2\%$ across the entire time range. However, the fit is not exact, particularly at low $q$, due to the fact that $q_1$ is only a pseudo-isosbestic point. This demonstrates that the apparent success of the model should be interpreted with caution, as it reflects an effective structural evolution rather than the transformation of one species into another. It is a consequence of the strong 
conservation constraints -- the Porod invariant and the geometric invariance of 
$q_2$ -- that confine the redistribution of scattering intensity to a 
quasi-two-state pattern, even as the underlying structural evolution remains 
continuous and hierarchical.

Having established the limits of the two-state interpretation of gelation, we now turn to the physical origin of the two isosbestic points. We first consider the high-$q$ isosbestic point, whose origin can be understood from local particle packing. We then show that the low-$q$ pseudo-isosbestic point arises from a different mechanism associated with the redistribution of scattering intensity over larger length scales.


\subsection{The high-$q$ isosbestic point ($q_2,I_2$)}
\begin{figure*}
\centering
\includegraphics[width=1\textwidth]{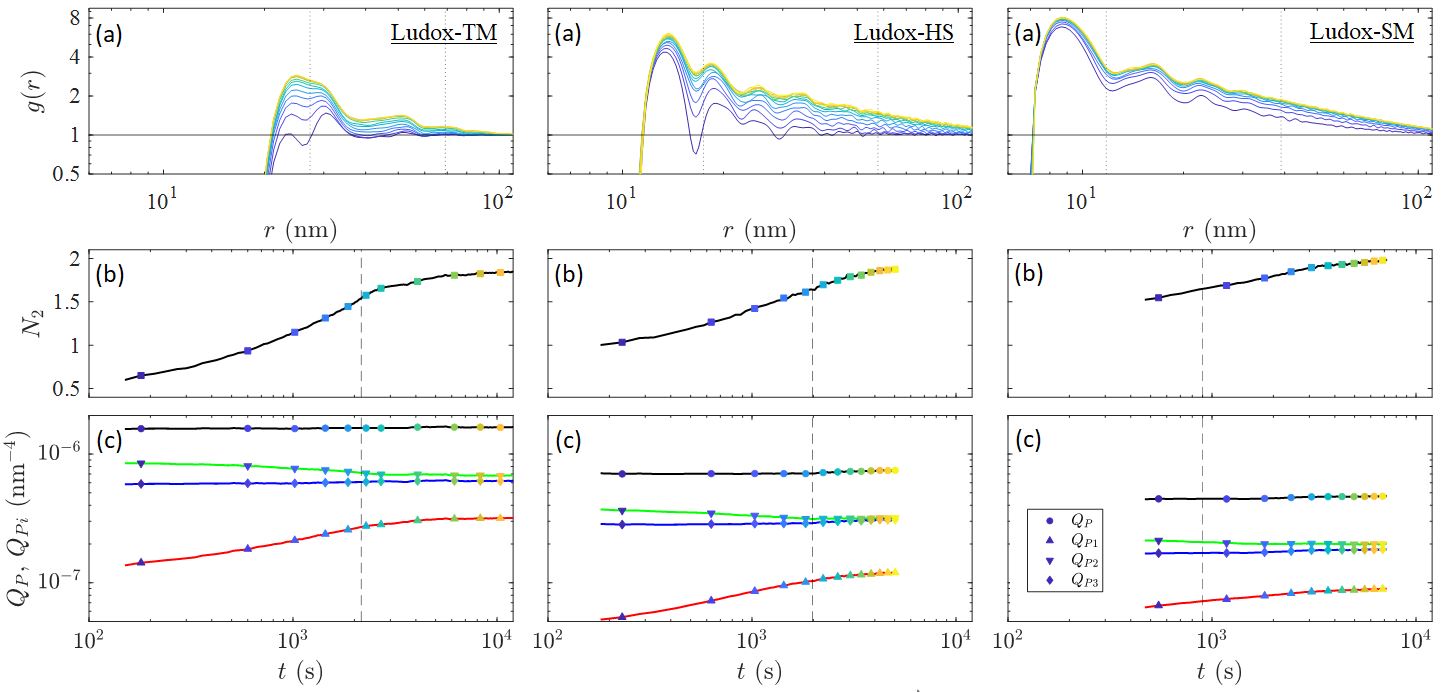}
    \caption{(a) Time evolution of the radial distribution function $g(r)$ a function of the distance $r$ between two colloids center of mass. $g(r)$ is calculated from $S(q)$ as shown in Fig.~\ref{fig:iq}(d). 
    (b) Time evolution of the number of particle in the first neighbor shell, up to $r_2=2\pi/q_2$ calculated from $g(r)$.
    (c) Time evolution of the porod invariant $Q_p$ and the sectioned ones $Q_{pi}$. 
    }
    \label{fig:gr}
\end{figure*}

The high-$q$ isosbestic point $q_2$ corresponds to a length scale on the order of a single particle $N(q_2)=\pi/(q_2r_0)\simeq 1$ (Fig.~\ref{fig:iso} (c)) and is associated with local, short-range correlations. To better understand its meaning, let us focus on the time evolution of the radial distribution function $g(r)$. For an isotropic system, $g(r)$ can be extracted from the structure factor $S(q)$.

For ludox dispersions, the time evolution of $g(r)$ is plotted in Fig.~\ref{fig:gr}(a).
The radial distribution function is zero for $r < 0.9\times 2r_0$ reflecting  the volume-exclusion interactions between particles. This distance is slightly smaller than $2r_0$, due to the polydispersity of the particles. 
At larger separations, $g(r)$ increases dramatically and exhibits a sequence of alternating maxima and minima whose amplitudes progressively decay toward unity as $r$ increases. Each successive minimum defines the boundaries of the first, second, third, etc., coordination shells. This damped oscillatory behavior reflects short-range structural ordering, which gradually disappears at large distances as spatial correlations vanish and $g(r) \to 1$, indicating that the medium becomes homogeneous on large length scales.

The high-$q$ isosbestic point corresponds, in real space, to $r_2 = 2\pi/q_2$. This distance coincides with the first minimum of $g(r)$. We note that this minimum is time-invariant, indicating that $r_2$ consistently marks the outer boundary of the first coordination shell at all times. Within this first-neighbor shell, both the height of the first maximum and the depth of the adjacent minimum increase over time. This evolution reflects a progressive strengthening of short-range correlations and an increase in the average number of nearest neighbors as gelation proceeds. Quantitatively, this effect is captured by the coordination number, $N_c = 4\pi n \int_0^{r_\mathrm{cut}} r^2 g(r)\text{d}r$, where $n = \phi / V_\mathrm{part}$ is the particle number density and $r_\mathrm{cut} = r_2$. As shown in Fig.~\ref{fig:gr}(b), $N_c$ increases with time, from $N_c < 1$ at short 
times to $N_c \approx 2$ at long times, demonstrating the progressive buildup of local connectivity within the forming gel network. This value is compatible with fractal gels formed at low volume fraction and strong interparticle attraction~\cite{rouwhorst2020}. It is worth noting that $N_c \simeq 2$ does not imply a purely linear structure: branching nodes ($z = 3$) and dangling ends ($z = 1$) can coexist in significant numbers without altering the average 
coordination. In a percolated fractal gel, most particles are incorporated into the space-spanning network, although small isolated aggregates may persist and contribute to slightly lower values of $N_c$. The value $N_c \simeq 2$ therefore reflects the low overall connectivity of the network backbone rather than its detailed local topology, and is indeed consistent with a structure built from linear chains, ramified chains, and chains with dangling ends.

In summary, the isosbestic point $q_2$ corresponds to a characteristic geometric scale associated with particle contact induced by attractive interactions. In real space, this length scale is related to the first coordination shell in $g(r)$. For $t \gg t_{\mathrm{gel}}$, we measure a coordination number $N_c \simeq 2$, consistent with the formation of a percolating space-spanning network composed of linear chains with  branching nodes and dangling arms.


\subsection{The low-$q$ isosbestic point ($q_1,I_1$)}
For the ludox dispersions, the low-$q$ isosbestic point $q_1$ reflects correlations at a larger, mesoscopic length scale corresponding to the emerging gel network structure. Unlike the high-$q$ isosbestic point $q_2$, which marks a crossover between particle-scale and mesoscopic regimes, $q_1$ serves as a pivot separating different modes of structural rearrangement during gelation: scattering intensity is transferred from higher $q$ (particle and local aggregate scale) to lower $q$ (network and fractal cluster scale) as the gel forms, while the intensity at $q_1$ itself remains invariant throughout the sol--gel transition. 

In a first interpretation, the wavevector $q_1$ can be associated with a characteristic mesoscopic length scale defined as $\ell_{\mathrm{meso}} \sim 2\pi/q_1$. Within this geometric picture, the number of particles spanning the thickness of a typical gel strand can be estimated as $n_{\mathrm{strand}} \sim \pi/(q_1 r_0)$ and range from 3 to 4 particles depending on the gel as shown in Fig.~\ref{fig:iso}(c). 
This interpretation, however, must be treated with caution. It assumes a direct and unique mapping between a single reciprocal-space position and a well-defined real-space strand thickness, which is an oversimplification for disordered gels. In reality, the scattering intensity at a given $q$ reflects contributions from a broad distribution of structural environments and length scales. Gel strands may exhibit heterogeneity in thickness, branching, and internal restructuring, none of which can be captured by a single characteristic length extracted from $q_1$. A rigorous validation of this picture would therefore require real-space analysis or complementary imaging techniques.

Alternatively, the emergence of isosbestic points in the SAXS intensity can be interpreted in terms of conservation of the scattering invariant. As shown in Fig.~\ref{fig:gr}(c), the Porod invariant $\mathcal{Q}_P = \int_0^{\infty} q^2 I(q) \text{d}q = 2\pi^2 (r_e \Delta\rho)^2 \phi (1-\phi)$
remains constant throughout the sol--gel transition, as it depends only on the scattering contrast $r_e\Delta\rho$ and the particle volume fraction $\phi$. Importantly, this relation follows from a two-phase decomposition of the sample (colloids and solvent) and from Parseval's theorem alone (Appendix~B): it makes no assumption about particle interactions, spatial arrangement, or the specific form of $S(q)$, and therefore remains valid irrespective of the restructuring, clustering, and network formation occurring during gelation. Its two underlying requirements --- a sharp, two-level particle--solvent interface, and conservation of the bulk composition $(\phi,\Delta\rho)$ --- are independently confirmed in our data: the high-$q$ intensity follows a clean $I(q)\propto q^{-4}$ Porod law with a time-invariant prefactor, consistent with a sharp interface, while $\mathcal{Q}_P$ itself is constant in time, which is a non-trivial check since any process altering the effective two-phase composition (solvent expulsion, particle swelling, or a change in effective contrast due to interfacial hydration) would necessarily cause it to drift. From the measured invariant, we obtain $\phi = \frac{1}{2}\left(1 - \sqrt{1 - 4\frac{\mathcal{Q}_P}{2\pi^2 (r_e \Delta\rho)^2}}\right)$, whose values are tabulated in Appendix.A Tab.~\ref{tab:ludox} and are in good agreement with the value independently derived from the weight concentration $c_w$, providing further support for the validity of the two-phase description.

In the Kratky representation (Fig.~\ref{fig:iq}(c)), which emphasizes the invariant through the $q^2 I(q)$ weighting, the structural evolution appears as a redistribution of scattering intensity: $q^2 I(q)$ increases at low $q$ ($q<q_1$) and decreases in the intermediate range ($q_1<q<q_2$). The constancy of $\mathcal{Q}_P$ therefore reflects a balance between these opposing contributions, rather than any breakdown of the underlying two-phase assumption.
To quantify this redistribution, the invariant was partitioned into three regions,
\begin{equation}
\begin{split}
\mathcal{Q}_P(t) &= 
\underbrace{\int_{\min(q)}^{q_1} q^2 I(q,t)\,\text{d}q}_{\mathcal{Q}_{P1}(t)}
+
\underbrace{\int_{q_1}^{q_2} q^2 I(q,t)\,\text{d}q}_{\mathcal{Q}_{P2}(t)} \\
&\quad +
\underbrace{\int_{q_2}^{\max(q)} q^2 I(q,t)\,\text{d}q}_{\mathcal{Q}_{P3}(t)}.
\end{split}
\end{equation}
As shown in Fig.~\ref{fig:gr}(c), $\mathcal{Q}_P$ and $\mathcal{Q}_3$ remain constant, while $\mathcal{Q}_1$ increases at the expense of $\mathcal{Q}_2$, demonstrating a transfer of scattering weight from local to mesoscopic length scales during gelation.


As detailed in Appendix.B, the Porod invariant is proportional to the variance of phase indicator in the system.
The phase indicator $\eta(\mathbf r)$ is a binary function that identifies the local phase of the system, taking the value $1$ if the position $\mathbf r$ lies inside a particle (with probability $\phi$) and $0$ otherwise.
More precisely, $\mathcal{Q}_p = 2\pi^2 (r_e\Delta \rho)^2  \mathrm{Var}[\eta]$. This remains true for the $\mathcal{Q}_{Pi}$ except it corresponds to a coarse grain variance associated with the length scale defined by the integral boundaries: $\mathcal{Q}_{pi} = 2\pi^2 (r_e\Delta \rho)^2  \mathrm{Var_i}[\eta]$ as detailed in Appendix.C. These reciprocal-space intervals correspond to real-space length-scale ranges
$[\ell_1] = 2\pi\left[\frac{1}{q_1},\,\frac{1}{\min(q)}\right]$, 
$[\ell_2] = 2\pi\left[\frac{1}{q_2},\,\frac{1}{q_1}\right]$,
$[\ell_3] = 2\pi\left[\frac{1}{\max(q)},\,\frac{1}{q_2}\right]$.
In this interpretation, the total variance of the phase indicator,
$\mathrm{Var}[\eta]=\phi(1-\phi)$ is constant over time  and is distributed among different structural length scales. Each term $\mathrm{Var}_i[\eta]$ quantifies the fraction of the total two–phase density fluctuations carried by structures whose characteristic sizes lie within the real–space interval associated with $q$-intervals \textcircled{1}, \textcircled{2} and \textcircled{2}.
In other words, $\mathcal{Q}_{P_i}$ provides a coarse-grained measure of the density fluctuations at the spatial scales associated with the interval $[\ell_i]$: the lower boundary of the interval sets the minimal size of structures contributing to that variance, while the upper boundary defines the maximal size of the contributing structures.
%
The high-$q$ invariant $\mathcal{Q}_{P_3}$, associated with short length scales below or near the particle diameter, remains essentially constant because it reflects the variance at the particle–solvent interfaces, which do not change during aggregation. 
The intermediate-$q$ invariant $\mathcal{Q}_{P_2}$ decreases over time. During gelation, particles become progressively trapped within the interaction potential of their neighbors (length scale $2\pi/q_2$) and concurrently small groups of 3 to 4 (length scale $2\pi/q_1$) neighboring particles form rigid clusters that dynamically freeze as the gel network develops. Zone \textcircled{2} is the region on which we monitor dynamical arrest. 
The growth of $\mathcal{Q}_{P_1}$ during gelation reflects the emergence of large-scale density fluctuations in the system. Several mechanisms contribute to this increase. First, clusters of locally arrested particles (3--4 particles) progressively merge into larger aggregates, enhancing the contrast at length scales corresponding to Zone~\textcircled{1}. Second, these clusters can still undergo thermal fluctuations as rigid units, which further contributes to density variations at long wavelengths. Third, the formation of a heterogeneous network of clusters and voids introduces additional spatial inhomogeneity. Collectively, these effects lead to a systematic increase of the low-$q$ invariant, indicating that $\mathcal{Q}_{P_1}$ captures the overall development of large-scale heterogeneities beyond the local particle cages probed by the intermediate-$q$ invariant $\mathcal{Q}_{P_2}$.

In summary, in this framework, the isosbestic point at $q_1$ acts as a natural pivot separating the intermediate-$q$ regime associated with locally arrested particle clusters from the low-$q$ regime that reflects large-scale density heterogeneities of size $\xi$ in the the emerging network.
Its invariance arises from the global constraint imposed by conservation of the Porod invariant
. It therefore provides a robust boundary for partitioning reciprocal space and quantifying the redistribution of structural correlations during network formation, independently of the detailed gelation pathway. 
Moreover, $q_1$ defines a characteristic coarse-graining length scale corresponding to clusters containing $N(q_1)$ particles. These clusters progressively undergo dynamical arrest and form the elementary building blocks from which the gel network can be segmented to build larger percolating clusters.

\begin{figure}
\centering
\includegraphics[width=0.471\textwidth]{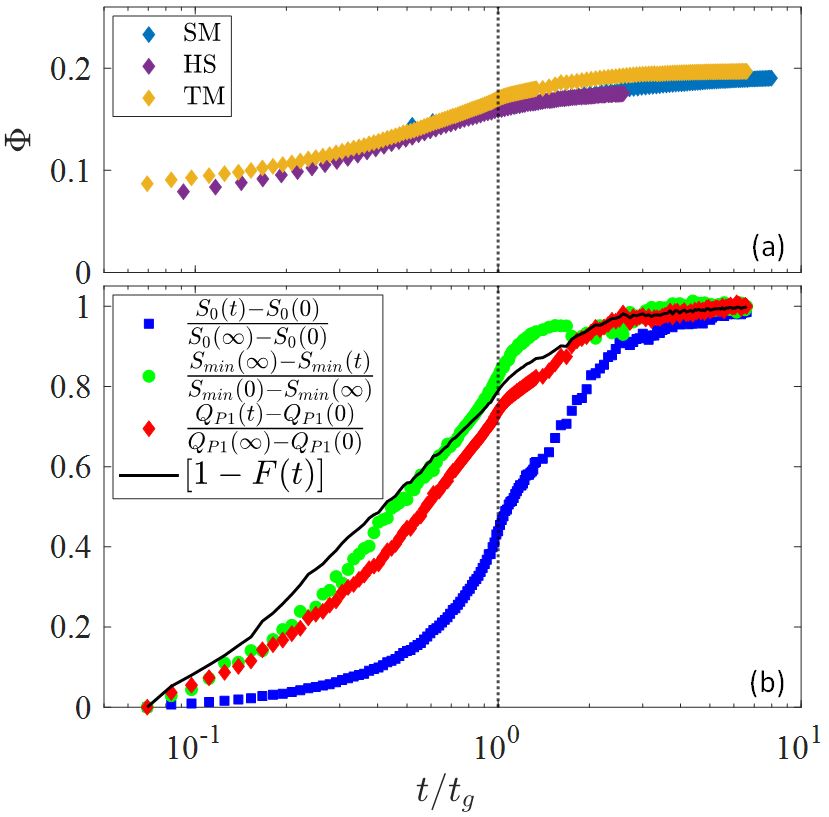}
    \caption{(a) Evolution of $\Phi$ as function of the normalized time $t/t_g$ for the 3 different ludox gels. (b) Evolution of $[1-F(t)]$ and the normalized $S_0(t)$, $S_{min}(t)$ and $\mathcal{Q}_{P1}(t)$ as a function of $t/t_g$ for the ludox-TM dispersion.
    }
    \label{fig:Phi}
\end{figure}

Based on the separation of the SAXS invariant, we define a dimensionless parameter
\begin{equation}
\Phi(t) = \frac{\mathcal{Q}_{P1}(t)}{\mathcal{Q}_P},
\end{equation}
which quantifies the fraction of the total scattering invariant carried by mesoscopic, network-scale structures. As shown in Fig.~\ref{fig:Phi}(a), $\Phi(t/t_{\rm g})$ exhibits a sigmoidal shape across all three Ludox systems: it rises from a small initial value, accelerates around $t_{\rm g}$, and plateaus at long times. While this qualitative behavior is universal, 
the precise rate of increase and plateau value are system-dependent, reflecting differences in the large-scale heterogeneity of the final network. Rather than collapsing onto a master curve, $\Phi(t/t_{\rm g})$ thus acts as a simple, model-free fingerprint of the gelation pathway.

\subsection{Probing gelation kinetics via $S(0)$ and  $S(q_{min})$}

So far, we have focused on the isosbestic points; however, the kinetics of gelation is also encoded in the evolution of $S_0(t)=S(q\rightarrow 0, t)$ and $S_{min}(t)=S(q_{min},t)$, where $q_{\rm min}$ corresponds to the position of the minimum of $S(q)$ for $q_1 < q_{\rm min} < q_2$ and is invariant with time. Dimensionless values are obtained by normalizing these quantities by their respective values at $t\to\infty $.  The time evolution of these two parameters is shown in Fig. ~\ref{fig:Phi}(b) for sample TM, for which the kinetics the full kinetics is captured. 
It is evident that the dynamics of $S_0(t)$ and  $S_{min}(t)$ differ markedly: $S(q_{min})$ evolves much faster toward saturation than $S_0(t)$. 
In particular, $S_{min}(t)$ rises steeply to a plateau at a time corresponding to the gelation time  $t_{g}$ defined by the intersection of $G'$ and $G''$, whereas $S_0(t)$ approaches its saturation value much more gradually. 
These distinct behaviors indicate that the two parameters, measured at different $q$ positions, carry complementary information. The deepening of the minimum structure factor between $q_1<q<q_2$ probes the formation of typical length strands  $3.5R$ (at $q=q_{min}$), whereas $S_0(t)$ reflects the evolution of large-scale clusters. The rapid deepening of $S_{min}(t)$ signals fast local structuring of aggregates, corresponding to early formation of the backbone (strands) and possibly to local densification or internal organization of the clusters. 
By contrast, slower growth of $S_0(t)$ indicates that local long-range correlations are gradual: percolation and coalescence of structures take time, and existing clusters continue to connect over increasingly large length scales. The combined behavior of $S_0(t)$ and  $S_{min}(t)$ suggests a two-step hierarchical assembly: rapid formation of locally dense strands and clusters, followed by slower coalescence and long-range percolation of the gel network. This reflects both the mesoscopic structuring of the backbone and the progressive development of network connectivity, which together control the mechanical and structural of the gel properties.

\section{Discussion}

The results presented above demonstrate that isosbestic points in time-resolved SAXS carry precise structural information during colloidal gelation, but that their physical origin differs fundamentally from the two-state interconversion picture borrowed from absorption spectroscopy. In this section, we place these findings in a broader context by comparing them with other systems where isosbestic or iso-scattering points have been reported, by discussing the distinct characters of $q_1$ and $q_2$. We finally discuss the role model free analysis and examine the significance of $\Phi(t/t_g)$ as a descriptor of gelation.

\subsection{A taxonomy of isosbestic points in scattering}

The observation of invariant crossing points in scattering experiments has been reported across a wide range of systems and physical mechanisms. Rather than treating each case in isolation, it is instructive to classify them according to their physical origin, which we argue falls into two distinct categories.

The first category corresponds to systems where the spectroscopic analogy is genuinely valid and the scattering intensity can be written as a linear superposition of two distinct structural contributions as in Eq.~\ref{eq:2state}. The study by Nicolai \emph{et al.}~\cite{nicolai2006} on heat-induced aggregation of globular proteins is the clearest example, with isosbestic points arising from a well-defined two-state conversion between native and denatured proteins. Similarly, Goodell \emph{et al.}~\cite{goodell2008} reported iso-scattering points during water vapor-driven crystallization of nanoparticles, consistent with two-state coexistence between amorphous and crystalline populations, and Tyburski and Nilsson~\cite{tyburski2024} identified an analogous feature in equilibrium structural fluctuations of liquid water, interpreted in terms of two interconverting local hydrogen-bond configurations. Sauter \emph{et al.}~\cite{sauter2016} provide a more subtle example: isosbestic points observed during a temperature ramp in metastable $\beta$-lactoglobulin aggregates could be well described by the linear superposition model, yet the two states correspond not to two coexisting thermodynamic phases but to two internal structural configurations of the same aggregate --- a continuous internal reconfiguration within pre-existing protein clusters, from a high-temperature state where monomers retain local mobility 
to a low-temperature state where the dimer locks in as the elementary building block. In all cases belonging to this first category, the two-state superposition model provides a valid mathematical and physical description, even if the nature of the two states varies considerably from one system to another.
We note that the two-state model carries strong assumptions.  The linear superposition therefore requires specific conditions, such as statistical independence or spatial separation of the contributing structures so that cross scattering term from the two potulation can be negleted. These conditions are not fulfilled in the case of Ludox gelation. First, the system undergoes a continuous structural evolution, in which particles progressively aggregate and reorganize into a network, rather than transitioning between two well-defined populations. Second, the growing clusters remain spatially correlated and interpenetrating, so that cross-interference terms cannot be neglected. As a result, the assumption of a linear superposition of two independent scattering states is not physically justified, which explains the failure of the simple two-state model (Eq.\ref{eq:2state}) to describe the observed kinetics.

The second category, corresponding to the system studied here, belongs to a class of transitions marked by continuous structural evolution rather than discrete interconversion between species. In salt-induced colloidal gelation, there are no distinct ``unaggregated'' and ``aggregated'' populations with time-invariant scattering functions; instead, $S(q,t)$ evolves continuously as particles diffuse, aggregate, and percolate into a space-spanning network, while the form factor $P(q)$ remains unchanged throughout. The full two-state model, including the cross-interference term of Eq.~\ref{eq:2state_b}, captures this evolution remarkably well, although it cannot fully reproduce the small but systematic drift observed in the low-$q$ crossing point. This crossing point is therefore only pseudo-isosbestic rather than a true fixed point, and its drift constitutes a hallmark of this type of transition, reflecting the gradual redistribution of large-scale interparticle correlations during network formation rather than interconversion between two invariant structural states. The appearance of a similar two-isosbestic-point pattern in covalently cross-linked protein gels reported by Kaieda \emph{et al.}~\cite{kaieda2014} --- a system with a fundamentally different bonding mechanism --- suggests that this pattern is a generic signature of gelation as a continuous process rather than a peculiarity specific to the colloidal system examined here.

\subsection{The distinct characters of $q_2$ and $q_1$}

Although $q_1$ and $q_2$ both qualify as isosbestic points, their physical origins are fundamentally different and it is important not to treat them symmetrically.
The high-$q$ isosbestic point $q_2$ has a geometric, single-particle origin. Its position is set by the condition that $r_2 = 2\pi/q_2$ coincides with the first minimum of $g(r)$, i.e. the outer boundary of the first coordination shell. This distance is imposed by hard-core repulsion and attractive contact between neighboring particles, and is therefore determined primarily by the particle radius $r_0$. As a 
consequence, $q_2 r_0 \approx \pi$ is approximately constant across all three Ludox systems (Appendix.A Tab.~\ref{tab:ludox}), and the value $q_2 r_0 \approx 1.2$ reported by Kaieda \emph{et al.}~\cite{kaieda2014} for cross-linked protein gels is consistent with this geometric picture. The universality of $q_2$ in reduced coordinates reflects the fact that particle contact is a universal geometric constraint, 
independent of the specific interaction potential or bonding mechanism driving gelation. In real space, the invariance of $q_2$ signals that the outer boundary of the first coordination shell does not shift during gelation, even as the height of $g(r)$ at contact grows progressively and the coordination number $N_c$ increases from below unity toward $N_c \simeq 2$.

\textcolor{black}{The interpretation of $q_2$ as a local geometric marker should also be distinguished from a full topological characterization of the gel backbone. The increase of the coordination number extracted from $g(r)$ demonstrates the buildup of local contacts, but the average coordination alone cannot determine whether those contacts belong to dangling ends, locally ramified clusters, loops, or load-bearing strands. This distinction is important because previous network-based analyses of colloidal gels have shown that mechanically relevant connectivity is encoded not only in the number of contacts, but also in how those contacts are organized into resilient, rigid, or stress-bearing substructures. Thus, $q_2$ is best viewed as a robust marker of local contact formation and first-shell organization, while additional real-space or network-level information would be required to determine how those contacts contribute to elasticity or yielding.}

The low-$q$ isosbestic point $q_1$ has a collective, kinetics origin that is qualitatively different. Its invariance does not reflect a fixed geometric length scale but rather arises as a consequence of the conservation of the Porod invariant $Q_P = 2\pi^2(r_e\Delta\rho)^2\phi(1-\phi)$, which is time-independent since neither the contrast $\Delta\rho$ nor the volume fraction $\phi$ changes during gelation. As scattering weight is transferred from the intermediate-$q$ regime toward low $q$, $q_1$ emerges as the natural pivot wavevector at which these opposing contributions balance. Its position in reduced coordinates $N(q_1) = \pi/(q_1 r_0)$ is \emph{not} universal: it varies between systems, taking values between 3 and 4 across the three Ludox dispersions, reflecting differences in the size of the elementary arrested building blocks set by the specific gelation conditions (volume fraction, salt concentration, interaction potential). However, its \emph{physical meaning} is general: $q_1$ always marks the coarse-graining length scale at which dynamical arrest first occurs, separating the intermediate-$q$ regime of locally frozen clusters from the low-$q$ regime of large-scale network heterogeneities. In this sense, $q_1$ is a system-specific quantity that nevertheless carries a universal structural interpretation.

\textcolor{black}{The collective nature of $q_1$ makes it particularly interesting from the perspective of network formation. Unlike $q_2$, which is fixed by near-contact geometry, $q_1$ reflects the redistribution of density fluctuations across intermediate and large length scales. It therefore resembles a coarse-graining boundary rather than a direct measurement of a unique strand diameter. This distinction is important because colloidal gels generally contain heterogeneous strands, branch points, dangling arms, and voids, so a single wavevector cannot be mapped one-to-one onto a single real-space object. Instead, $q_1$ should be interpreted as the reciprocal-space scale at which the system transfers structural weight from locally arrested clusters to larger network heterogeneities. In this sense, $q_1$ provides a model-free experimental analogue of a network coarse-graining scale: it identifies the length scale at which local aggregates cease to behave as independent objects and begin to contribute to the growing mesoscopic gel network.}

\subsection{A model-free framework for colloidal gelation}

A key strength of the analysis presented in this work is that it is entirely model-free and can be applied to any system exhibiting isosbestic points in time-resolved SAXS, irrespective of their physical origin or the category to which they belong in the taxonomy introduced above.

This approach stands in sharp contrast to, yet is complementary with, conventional approaches to colloidal gel analysis, which typically require fitting $I(q,t)$ to a fractal scattering model such as those developed in~\cite{freltoft1986,teixeira1988,beaucage1995} to extract parameters such as the fractal dimension $d_f$, the correlation length $\xi$, and the cluster size. While such fits provide valuable physical insight into the geometry and morphology of the gel network, they are sensitive to the choice of model, the $q$-range used, and the assumed form of the structure factor, and they become increasingly ambiguous during the dynamic process of gelation where multiple structural length scales evolve simultaneously. The present framework sidesteps these difficulties by working directly with conserved quantities and geometric constraints, providing an unambiguous and assumption-free characterisation of the structural evolution.

\textcolor{black}{The present framework should be viewed as complementary to particle-resolved and network-based approaches rather than as a replacement for them. Scattering provides robust ensemble-averaged information about how density correlations are redistributed across length scales, while network analysis can identify which subsets of contacts, clusters, or strands are mechanically active. This distinction is central in colloidal gel rheology: two structures may exhibit similar cluster sizes, coordination numbers, or scattering features, yet differ in their load-bearing backbones, bond-age distributions, or yielding pathways. The value of the isosbestic analysis is therefore that it supplies experimentally accessible, model-free markers of when structural weight moves from local to mesoscopic scales. These markers can then be compared, in future work, with real-space network descriptors such as percolation, backbone fraction, dangling-end population, rigidity, or stress localization.}

This model-free character confers several practical advantages for at least two reasons. \textit{(i)} It makes the analysis robust to the complexity and heterogeneity of real gel networks, where the spatial organization of clusters, strands, and voids defies simple analytical models.
\textit{(ii)} This framework is computationally lightweight --- extracting $q_1$, $q_2$, $\Phi(t)$, and the Porod invariant decomposition requires only numerical integration of the measured $I(q,t)$ --- which makes it well suited for high-throughput SAXS screening of colloidal 
formulations, where rapid, assumption-free structural diagnostics are particularly valuable.

\subsection{$\Phi(t/t_g)$ as a fingerprint of gelation}

The decomposition of the Porod invariant into partial contributions $Q_{P1}$, $Q_{P2}$, and $Q_{P3}$, using $q_1$ and $q_2$ as natural boundaries, provides a scale-resolved picture of structural reorganization during gelation that goes beyond what can be extracted from the scattering profiles directly. The dimensionless parameter $\Phi(t) = Q_{P1}(t)/Q_P$, which quantifies the fraction of total scattering variance carried by structures at mesoscopic, network-scale length scales, offers a particularly compact and informative descriptor of this process.

Several features of $\Phi(t/t_g)$ are noteworthy. First, it is strictly model-free: no assumption about the functional form of $S(q,t)$, the fractal dimension, or the cluster size distribution is required to compute it. It follows directly from the measured $I(q,t)$ and the positions of the isosbestic points. Second, it compresses the full two-dimensional structural evolution encoded in $I(q,t)$ into a single 
scalar function of time, making it straightforward to compare across systems and experimental conditions. Third, as shown in Fig.~\ref{fig:Phi}, the sigmoidal shape of $\Phi(t/t_g)$ is qualitatively universal across the three Ludox systems despite differences in particle size, volume fraction, and salt concentration: $\Phi$ rises from a small initial value, accelerates around the gelation time $t_g$, and plateaus at long times. The precise rate of increase and the plateau value, however, are system-dependent, encoding information about the specific gelation pathway. In this sense, $\Phi(t/t_g)$ functions simultaneously as a qualitative universal marker of the sol--gel transition and as a quantitative, system-specific fingerprint.

\textcolor{black}{The interpretation of $\Phi(t)$ as a gelation fingerprint is particularly useful because it separates a structural transition from a specific microscopic model. However, $\Phi(t)$ should not be interpreted as a direct measure of mechanical connectivity by itself. Because $\Phi$ measures the fraction of the scattering invariant carried by low-\textit{q}, mesoscopic structures, it is sensitive to the growth of density heterogeneity and network-scale correlations. The emergence of elasticity, by contrast, depends on whether those correlations form a mechanically connected and load-bearing backbone. This distinction is consistent with recent studies showing that colloidal gel rheology can depend on bond dynamics, network topology, and formation history in ways that are not captured by static structural descriptors alone. Thus, $\Phi(t)$, $S(q_{min},t)$, and $S(0,t)$ should be regarded as a minimal reciprocal-space set of observables that identify the timing and scale of structural reorganization, while their connection to modulus, yielding, or aging requires comparison with mechanical or network-level measures.}

The plateau value of $\Phi$ deserves particular attention. It reflects the fraction of total scattering variance that has been transferred to mesoscopic length scales once the gel is fully formed. A higher plateau indicates a more heterogeneous gel network with larger density fluctuations at long wavelengths, while a lower plateau corresponds to a more homogeneous structure. 
The differences in plateau values across the three Ludox systems is quite small suggesting that the large-scale heterogeneity of the final network are similar in the ludox gels tested.

The analysis of $S_0(t)$ and $S(q_{\min},t)$ provides complementary information to that captured by $\Phi(t)$. While $\Phi(t)$ offers a global, scale-integrated measure of how scattering intensity redistributes between different $q$-regions, $S_0(t)$ and $S(q_{\min},t)$ probe the gelation kinetics at specific, physically meaningful length scales. As shown in Fig.~\ref{fig:Phi}(b), the normalized evolution of $S(q_{\min},t)$ tracks the formation and densification of local structures at the scale of interparticle correlations, and coincides remarkably well with both the normalized evolution of $\mathcal{Q}_{P1}(t)$ and $1-F(t)$. This convergence of three independently defined observables onto a single kinetic curve is a strong internal consistency check, confirming that the intermediate-$q$ regime captures a well-defined structural process: the progressive arrest of local particle clusters into the gel backbone. By contrast, $S_0(t)$ evolves on a distinctly slower timescale, reflecting the gradual buildup of long-range density fluctuations as clusters coalesce and the network becomes globally connected. This decoupling between fast local structuring and slower large-scale organization -- not directly accessible from $\Phi(t)$ alone -- is a direct, model-free signature of the hierarchical nature of gelation, and mirrors the two-step assembly picture that emerges from the Porod invariant decomposition. Together, $\Phi(t)$, $S(q_{\min},t)$, and $S_0(t)$ form a minimal yet complete set of observables for characterizing gelation kinetics across scales, without any assumption about the functional form of the structure factor or the geometry of the network.

\subsection{Limitations and perspectives}

Several assumptions underlying the present analysis deserve discussion. The interpretation of the isosbestic points in terms of $S(q,t)$ evolution relies on the constancy of the form factor $P(q)$ throughout gelation, which holds for rigid inorganic colloids such as Ludox but may break down in systems where particles deform, swell, or undergo internal structural changes during aggregation. In such cases, the isosbestic points would reflect a combination of $P(q)$ and $S(q,t)$ evolution, complicating their interpretation. The framework developed here would therefore need to be adapted before being applied to soft or deformable particles, protein aggregates, or emulsion droplets.

The estimation of $Q_P$ from experimental data also requires care, as it involves integrating $q^2 I(q)$ over the full $q$-range, which is never fully accessible in practice. As discussed in the Appendix.B, the low-$q$ and high-$q$ tails must contribute negligibly for the estimate to be reliable. In the present experiments, the SAXS window at ID02 is sufficiently broad that this condition is satisfied to 
within a few percent, as confirmed by the time-independence of the measured $Q_P$. However, for systems with very large correlation lengths or very small particles, the accessible $q$-window may be insufficient and systematic corrections would be required.

Finally, we note that the present analysis was carried out under quiescent conditions. Extending this framework to systems driven by external stimuli --- such as shear flow or temperature-dependent interaction potentials~\cite{gibaud2012, sauter2016} --- would reveal how the redistribution of scattering variance across length scales is modified by non-equilibrium conditions, and whether the universality of $\Phi(t/t_g)$ observed here persists under external forcing. Such experiments would also allow one to test whether the isosbestic points themselves shift under shear, which would provide direct evidence for shear-induced modifications of the elementary building block size 
or the contact geometry. \textcolor{black}{A natural extension would be to apply the same invariant-based analysis to systems whose formation pathway or deformation history is deliberately varied. Sequentially gelled or multicomponent colloidal networks provide a useful test case because the same particle interactions and compositions can produce different mesoscale morphologies depending on when each component is allowed to gel. In such systems, one could ask whether $q_1$, $q_2$, and $\Phi(t)$ are controlled primarily by instantaneous structure, by the pathway through which that structure formed, or by the emergence of mechanically active interspecies contacts. Similarly, under shear, one could examine whether the isosbestic points remain invariant, shift anisotropically, or disappear as the gel network undergoes bond breaking, densification, fragmentation, and restructuring. Such measurements would directly test whether the model-free SAXS markers identified here are universal signatures of gelation or whether they encode pathway-specific information about network formation and deformation.}

\section{Conclusion}

In this work, we investigate the origin and physical meaning of isosbestic points observed in time-resolved SAXS during salt-induced gelation of Ludox colloidal dispersions. Combining scattering and rheology across different particle sizes, we identify two robust isosbestic points, $q_1$ and  $q_2$ , which partition reciprocal space into three distinct structural regimes.

Unlike the classical two-state interpretation from absorption spectroscopy, these isosbestic points arise from continuous structural evolution. The high-$q$ point $q_2$ is a geometric marker related to the first coordination shell and reflects the growth of local connectivity. In contrast, the low-$q$  point $q_1$ is only a pseudo-isosbestic point and  originates from the conservation of the Porod invariant and captures the transfer of scattering intensity from intermediate to large length scales during network formation. This redistribution is quantified by a dimensionless parameter  $\Phi(t)$, which exhibits a universal sigmoidal evolution across systems while retaining sensitivity to specific gelation pathways. Overall, isosbestic and pseudo-isosbestic points emerge as robust structural markers linking local particle organization to large-scale network formation. The proposed framework provides a simple, model-free approach to characterize gelation kinetics and can be extended to a wide range of soft matter systems and non-equilibrium conditions. In addition, the combined temporal evolution of $S(q_{min},t)$  and $S(0,t)$ provides a direct, scale-resolved signature of gelation, respectively revealing a decoupling between rapid local structuring and the slower buildup of global network connectivity.

Finally, Beyond colloidal gelation, the framework introduced here -- based on conserved quantities, geometric invariants, and model-free decomposition of the scattering invariant -- is broadly applicable to any soft matter system in which time-resolved scattering can be measured during structural evolution. Immediate extensions include protein aggregation and gelation, clay suspensions, block copolymer ordering, and nanoparticle crystallization, all of which exhibit isosbestic-like features whose physical origin has remained unclear. More broadly, extending this analysis to systems driven by external stimuli -- such as shear flow, electric fields, or temperature-dependent interaction potentials~\cite{gibaud2012} -- would reveal how the redistribution of structural correlations across length scales is modified under non-equilibrium forcing, and whether the universal sigmoidal shape of $\Phi(t/t_g)$ persists or breaks down when the gelation pathway is externally controlled. Such experiments would provide a stringent test of the generality of isosbestic points as structural markers, and could establish the present framework as a standard 
diagnostic tool for high-throughput scattering studies of self-assembly across soft 
and biological matter.

\section*{authors contribution}
WS and TG carried out the experiments. AG and TG designed the project, analyzed the data and wrote the article. SJ contributed to the discussion. 

\section*{Data availability}
The data supporting this article, including the raw and reduced SAXS intensity profiles $I(q,t)$ are available from the corresponding author upon reasonable request.

\section*{Acknowledgements}
The authors are especially grateful to the ESRF for beamtime at the beamline ID02 (proposal SC-5099). We thank Lauren Matthews and Theyencheri Narayanan for helping us with experiment at the beamline ID02 and for discussions. 
This work was supported by the grants: ANR-17-CE07-0040 and FWF/ANR (FWF: 10.55776/PIN9311124; ANR-24-CE91-0012) as well as the European Union’s Horizon Europe Framework Program HORIZON under the Marie Skłodowska-Curie Grant Agreement 101120301 and the LabEx iMUST of the University of Lyon (ANR-10-LABX-0064). This work benefited from meetings within the French working group GDR CNRS 2019 `Solliciter LA Matière Molle' (SLAMM). 

%

\clearpage
\section*{Appendix}

\subsection{Summary of the ludox dispersion properties}

\begin{figure*}
\centering
\includegraphics[width=0.971\textwidth]{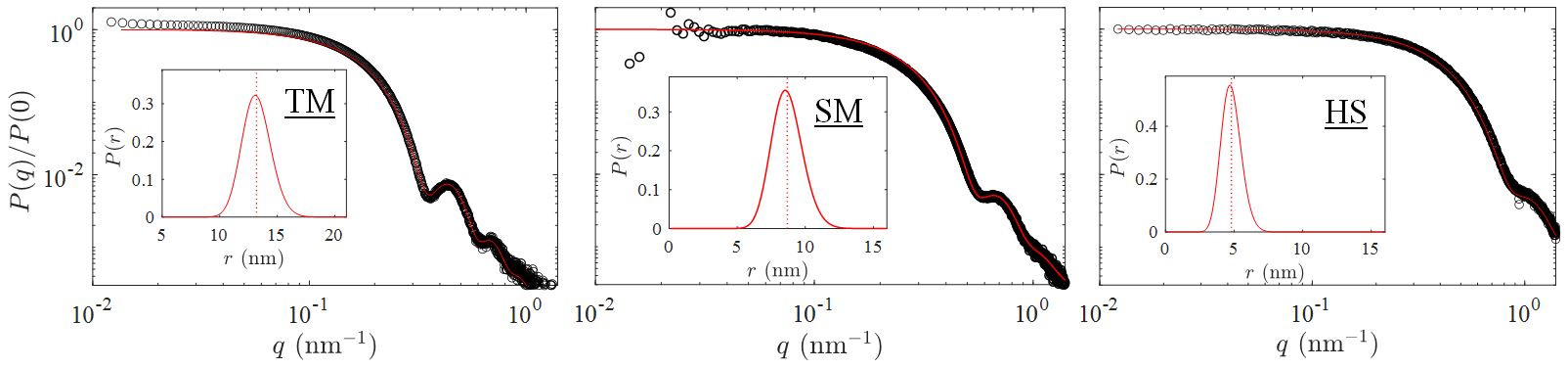}
    \caption{Measured form factor of the Ludox particles (symbols) and best fit using a polydisperse sphere model (red line). Inset: corresponding probability distribution of particle radii.
    }
    \label{fig:P}
\end{figure*}

Properties of Ludox dispersions for the three grades used in this study are summarized in  Tab.~\ref{tab:ludox}. The measure form factor $P(q)$ and their fit are displayed in Fig.~\ref{fig:P}.

\begin{table*}
\centering
\begin{tabular}{l|l|c|c|c}
\hline
Parameter & Description & SM & HS & TM \\
\hline\hline
$c_w$ & Weight concentration & 0.045 & 0.1 & 0.2 \\
$[\text{NaCl}]$ (mM) & Salt concentration & 500 & 610 & 700 \\
$\phi$ (\%) & Volume fraction from $c_w$ & 2.7 & 4.8 & 11.1 \\
$\phi_Q$ (\%)  & Volume fraction from invariant & 2.8 & 4.6 & 10.7 \\
$r_0$ (nm) & Particle radius from form factor & 4.70 & 7.78 & 12.15 \\
$dr_0/r_0$  & Relative polydispersity & 0.09 & 0.12 & 0.15 \\
$t_g$ (s) & Gelation time ($G'(t_g)=G''(t_g)$)  & 900 & 1980 & 2160 \\
$q_1$ (nm$^{-1}$) & Low-$q$ isosbestic point & $0.161\pm 0.006$ & $0.109\pm 0.005$ & $0.088\pm 0.005$ \\
$q_2$ (nm$^{-1}$) & High-$q$ isosbestic point &  $0.534\pm 0.005$ & $0.361 \pm 0.009$ & $0.242 \pm 0.002$ \\
$S(q_2)$ &  & 1.20 & 0.92 & 0.60 \\
$q_{min}$ (nm$^{-1}$) & Wave vector for which $S(q)$ is minimum  & 0.339  & 0.232 & 0.152 \\

\hline
\end{tabular}
\caption{Properties of Ludox dispersions for the three grades used in this study.}
\label{tab:ludox}
\end{table*}

\subsection{The Porod Invariant}

In the following we derive the porad invariant ($\mathcal{Q}_p=\int_0^\infty q^2 I(q) dq$) calculation. The key point is that the invariant is related to the variance of the phase field indicator. 

\paragraph{Step 1: Phase indicator field and its variance --}

Consider a two-phase system (particles in solvent) where $\phi$ is the particle volume fraction of particle. The phase indicator is
\begin{equation}
\eta(\mathbf r)=
\begin{cases}
1 & \text{if $\mathbf{r}$ is inside a particle, probability  $\phi$} \\\\
0 & \text{if $\mathbf{r}$ is outside a particle, probability  $1-\phi$} \\
\end{cases}
\label{eq:phaseindic}
\end{equation}
The spatial average of $\eta$ over the sample volume $V$ is $\langle \eta \rangle
= \frac{1}{V}\int_V \eta(\mathbf r)\,d^3r
= \phi.
$

The variance is defined as
\begin{equation}
\mathrm{Var}[\eta]
=
\langle (\eta-\langle\eta\rangle)^2 \rangle 
= \frac{1}{V} \int_V [\eta(\mathbf{r}) - \langle\eta\rangle]^2 \, d^3 r
\label{eq:vardef}
\end{equation}
and it is equal to the integral over the volume of flutuations square of of $\eta$. We denote these fluctuations $\delta \eta(\mathbf{r}) =\eta(\mathbf{r}) - \langle\eta\rangle$. The demonstration now proceeds by first expressing $\mathrm{Var}[\eta]$, in terms of the particle volume fraction $\phi$, and then relating it to the integrated scattering intensity.  

\paragraph{Step 2: Density fluctuations and volume fraction.}
The expectation value of any function $f(\eta)$ is $
\langle f(\eta) \rangle
=
\phi\, f(1) + (1-\phi)\, f(0).
$
Applying this definition to $f(\eta)=(\eta-\langle\eta\rangle)^2$ and using
$\langle\eta\rangle=\phi$ gives
\begin{equation}
\begin{aligned}
\mathrm{Var}[\eta] 
&= \langle (\eta-\phi)^2 \rangle \\
&= \phi (1-\phi)^2 + (1-\phi)(0-\phi)^2 \\
&= \phi (1-\phi)^2 + (1-\phi)\phi^2 \\
&= \phi(1-\phi)\left[(1-\phi)+\phi\right] \\
&= \phi(1-\phi)
\end{aligned}
\label{eq:phi}
\end{equation}

\paragraph{Step 2: Density fluctuations and scattering.}
The electron density can be written
\begin{equation}
\rho(\mathbf r)
=
\rho_s + \Delta\rho \, \eta(\mathbf r),
\end{equation}
where $\Delta\rho=\rho_p-\rho_s$ is the contrast between the particle and the background solvent. 
The fluctuations relative to the average density are
\begin{equation}
\delta\rho(\mathbf r)
=
\rho(\mathbf r)-\langle\rho\rangle
=
\Delta\rho [\eta(\mathbf r)-\langle\eta\rangle]=\Delta\rho\delta \eta(\mathbf r)
\end{equation}

The scattering amplitude is the Fourier transform of the density fluctuation $
A(\mathbf q)
=
r_e\int_V \delta\rho(\mathbf r)
e^{i\mathbf q\cdot\mathbf r}
d^3r$, where $r_e$ is the classical electron radius.
The scattered intensity is $I(\mathbf q)=|A(\mathbf q)|^2$/V.
Substituting $\delta\rho$, we get

\begin{equation}
I(\mathbf q)
=
\frac{(r_e \Delta\rho)^2}{V}
\left|
\int_V
\delta \eta(\mathbf r)
e^{i\mathbf q\cdot\mathbf r}
d^3r
\right|^2=
\frac{(\Delta\rho)^2}{V}
|\widetilde{\delta\eta}(\mathbf q)|^2.
\label{eq:powerspec}
\end{equation}

We obtain that $I(q)$ is the power spectrum of the phase indicator fluctuations.

\paragraph{Step 3: Parseval theorem.}

Parseval's theorem states that the total power of a function equals the total power of its Fourier transform:

\begin{equation}
\int_V |\delta\eta(\mathbf r)|^2 d^3r
=
\frac{1}{(2\pi)^3}
\int_{\mathbb R^3}
|\widetilde{\delta\eta}(\mathbf q)|^2
d^3q .
\end{equation}

Substituting the intensity expression (Eq.~\ref{eq:vardef}) and the variance expression (Eq.~\ref{eq:phi}), we obtain

\begin{equation}
\mathrm{Var}[\eta]
=
\frac{1}{(2\pi)^3(r_e\Delta\rho)^2}
\int_{\mathbb R^3} I(\mathbf q) d^3q .
\end{equation}

\paragraph{Step 4: Define the Porod invariant.}  

The integral over all $\mathbf{q}$-space of the scattering intensity defines the Porod invariant:
\begin{equation}
\mathcal{Q} \equiv \int_{\mathbb{R}^3} I(\mathbf{q}) \, d^3q.
\end{equation}

Using the variance of the phase indicator in Eq.\ref{eq:vardef},  we obtain
\begin{equation}
\boxed{
\mathcal{Q} = (2\pi)^3 (r_e\Delta \rho)^2  \mathrm{Var}[\eta] 
= (2\pi)^3 (r_e\Delta \rho)^2  \phi(1-\phi)
}.
\end{equation}

Note that for an isotropic system the Porod invariant simplifies to
\begin{equation}
\mathcal{Q} = 4\pi \int_0^\infty q^2 I(q) \, dq .
\end{equation}

Typically, the Porod invariant is redefined as 
$\mathcal{Q}_p = \mathcal{Q}/(4\pi)$ so that, for an isotropic system,
\begin{equation}
\boxed{
\mathcal{Q}_p = 2\pi^2 (r_e\Delta \rho)^2  \mathrm{Var}[\eta] 
= 2\pi^2 (r_e\Delta \rho)^2  \phi(1-\phi)
}.
\end{equation}

\paragraph{Practical considerations to estimate properly $\mathcal{Q}_p$.}

In practice, the Porod invariant $\mathcal{Q}_p=\int_0^\infty q^2 I(q) dq$ cannot be evaluated over the full range $q\in[0,\infty]$. Reliable estimates therefore require the experimental $q$-window to extend sufficiently toward both small and large wavevectors so that the missing contributions remain negligible. Two conditions must therefore be satisfied.
At low $q$, the intensity should approach a constant value, ensuring that sufficiently large length scales are probed such that the system appears homogeneous. In practice, it is sufficient that $I(q)\propto q^{\alpha}$ with $\alpha<2$, so that the integrand behaves as $q^2 I(q)\propto q^{2+\alpha}$ and vanishes as $q\rightarrow0$. The contribution from the unmeasured low-$q$ region is then negligible provided $q_{\min}$ is sufficiently small.
At high $q$, two-phase systems with sharp interfaces enter the Porod regime $I(q)\propto q^{-4}$. The integrand therefore decays as $q^2 I(q)\propto q^{-2}$, making the high-$q$ tail negligible once this regime is reached. In practice, this corresponds to measuring up to wavevectors where the structure factor approaches unity, $S(q_{\max})\simeq 1$.

\subsection{Decomposition of the Porod invariant.}

The Porod invariant can be partitioned into contributions arising from different regions of reciprocal space. In our case: 
\begin{equation}
\begin{split}
\mathcal{Q}_P(t) &= 
\underbrace{\int_{\min(q)}^{q_1} q^2 I(q,t)\,dq}_{\mathcal{Q}_{P1}(t)}
+
\underbrace{\int_{q_1}^{q_2} q^2 I(q,t)\,dq}_{\mathcal{Q}_{P2}(t)} \\
&\quad +
\underbrace{\int_{q_2}^{\max(q)} q^2 I(q,t)\,dq}_{\mathcal{Q}_{P3}(t)}.
\end{split}
\end{equation}

Using the relation between the Porod invariant and the variance of the phase indicator field, each partial invariant can be written as
\begin{equation}
\mathcal{Q}_{Pi} = 2\pi^2 (r_e\Delta\rho)^2 \, \mathrm{Var}_i[\eta],
\end{equation}
where
\begin{equation}
\mathrm{Var}_i[\eta] = \langle |\delta \eta_i(\mathbf r)|^2 \rangle
\end{equation}
represents the contribution to the variance of the phase indicator associated with density fluctuations at the spatial scales probed by the $q$-intervals defining the three regions:
\textcircled{1}~$[\min(q),\,q_1]$, 
\textcircled{2}~$[q_1,\,q_2]$, and 
\textcircled{3}~$[q_2,\,\max(q)]$. 
These reciprocal-space intervals correspond to real-space length-scale ranges
$[\ell_1] = 2\pi\left[\frac{1}{q_1},\,\frac{1}{\min(q)}\right]$, 
$[\ell_2] = 2\pi\left[\frac{1}{q_2},\,\frac{1}{q_1}\right]$,
$[\ell_3] = 2\pi\left[\frac{1}{\max(q)},\,\frac{1}{q_2}\right]$.
In this interpretation, the total variance of the phase indicator,
$\mathrm{Var}[\eta]=\phi(1-\phi)$, is distributed among different structural length scales. Each term $\mathrm{Var}_i[\eta]$ quantifies the fraction of the total two–phase density fluctuations carried by structures whose characteristic sizes lie within the real–space interval associated with $q$-intervals \textcircled{1}, \textcircled{2} and \textcircled{2}.
In other words, $\mathcal{Q}_{P_i}$ provides a coarse-grained measure of the density fluctuations at the spatial scales associated with $[\ell_i]$: the lower boundary of the interval sets the minimal size of structures contributing to that variance, while the upper boundary defines the maximal size of the contributing structures.

\end{document}